\documentclass{ifacconf}
\usepackage{graphicx}      
\usepackage{natbib}        

\usepackage{amsmath}
\usepackage{amssymb}
\usepackage{amsfonts}
\usepackage{graphicx}
\usepackage{lipsum}
\usepackage{epstopdf}
\usepackage{algorithmic}
\usepackage{mathptmx}

\ifpdf
  \DeclareGraphicsExtensions{.eps,.pdf,.png,.jpg}
\else
  \DeclareGraphicsExtensions{.eps}
\fi
\DeclareMathAlphabet{\bit}{OML}{cmm}{b}{it}

\def\fH{\mathfrak{H}}

\def\<{\leqslant}           
\def\>{\geqslant}           

\def\d{\partial}
\def\wh{\widehat}

\def\Re{\mathrm{Re}}   
\def\Im{\mathrm{Im}}   

\def\cH{\mathcal{H}}   

\def\mR{\mathbb{R}}    
\def\mC{\mathbb{C}}    

\def\Tr{\mathrm{Tr}}       
\def\rT{\mathrm{T}}        
\def\cov{\mathbf{cov}}

\def\eps{\epsilon}

\def\rank{\mathrm{rank}}       

\def\bE{\mathbf{E}}    
\def\bM{\mathbf{M}}    

\def\[[[{[\![\![}   
\def\]]]{]\!]\!]}   

\def\bra{{\langle}}
\def\ket{{\rangle}}

\def\rd{\mathrm{d}}        

\def\bJ{\mathbf{J}}

\def\x{\times}
\def\ox{\otimes}

\def\fF{\mathfrak{F}}

\def\sA{\mathsf{A}}
\def\sB{\mathsf{B}}
\def\sC{\mathsf{C}}
\def\sE{\mathsf{E}}
\def\sH{\mathsf{H}}

\def\sX{\mathsf{X}}

\def\cR{\mathcal{R}}
\def\cZ{\mathcal{Z}}
\def\cX{\mathcal{X}}

\def\cA{\mathcal{A}}
\def\cB{\mathcal{B}}

\def\mS{\mathbb{S}}

\def\ups{\upsilon}

\begin{document}
\begin{frontmatter}

\title{Measurement-based Initial Point Smoothing and Control Approach to Quantum Memory Systems\thanksref{footnoteinfo}}%
\thanks[footnoteinfo]{This work is supported by the Australian Research Council grant DP240101494.}
\vspace{-2mm}
\author[First]{Igor G. Vladimirov},\quad
\author[First]{Ian R. Petersen},\quad
\author[Second]{Guodong Shi}

\vspace{-2mm}
\address[First]{School of Engineering, Australian National University, Canberra, ACT,
Australia (e-mail: igor.g.vladimirov@gmail.com), i.r.petersen@gmail.com)}
\address[Second]{Australian Centre for Robotics, University of Sydney, Camperdown, Sydney, NSW, Australia (e-mail: guodong.shi@sydney.edu.au)}

\begin{abstract}                
This paper is concerned with a quantum memory system for storing quantum information in the form of its initial dynamic variables in the presence of environmental noise. In order to compensate for the deviation from the initial conditions,  the classical parameters of the system Hamiltonian are affected by the actuator output of a measurement-based classical controller. The latter uses an observation process produced by a measuring apparatus from the quantum output field of the memory system. The underlying system is modelled as an open quantum harmonic oscillator whose Heisenberg evolution is governed by linear Hudson-Parthasarathy quantum stochastic differential equations. The controller is organised as a classical linear time-varying system, so that the resulting closed-loop system has quantum and classical dynamic variables.
We apply linear-quadratic-Gaussian  control and fixed-point smoothing at the level of the first two moments  and   consider controllers with a separation structure which involve a continuously updated  estimate for the initial quantum variables. The initial-point smoother is used for actuator signal formation so as to minimise the sum of  a mean-square deviation of the quantum memory system variables at a given time horizon from their initial values and an integral quadratic penalty on the control signal.
\vspace{-3mm}
\end{abstract}

\vspace{-2mm}
\end{frontmatter}

\vspace{-2mm}
\section{Introduction}
\vspace{-3mm}

Unitary evolution of isolated quantum mechanical systems, specified by the system Hamiltonian  and algebraic structure of operator-valued dynamic variables, is a hallmark of quantum mechanics [\cite{S_1994}]. The reversibility  of such dynamics plays the role of a resource in quantum information processing and is employed,  for example, in the quantum gates as building blocks of quantum computation architectures [\cite{NC_2000}].
In the case of zero Hamiltonian (meaning the absence of internal interactions) and complete isolation  from the environment, a quantum system can serve as an ideal memory which preserves its initial quantum state and dynamic variables indefinitely.
However, coupling to  surroundings (such as external fields),  which is unavoidably present  in physical setups, makes the system drift away from the initial condition in an irreversible fashion. As a result, the ability to store quantum information may only be achieved to a limited extent and for a finite period of time.

The effects of external quantum noise [\cite{GZ_2004}]  on the system and its performance as a quantum memory can be modelled in the framework  of
open quantum dynamics by using the tools of quantum stochastic calculus  [\cite{HP_1984,P_1992}].   The quantum stochastic differential equations (QSDEs), which describe the   Heisenberg picture  evolution  of such a system and are driven by a quantum Wiener process as the environmental noise, can also involve the output fields of other quantum systems [\cite{GJ_2009}]. The resulting coherent (that is, measurement-free) interconnections through direct energy or field-mediated coupling of systems [\cite{ZJ_2011a}] or their measurement-based interconnections  (involving a measuring apparatus) [\cite{WM_2010}] are specified by individual and interaction Hamiltonians and coupling operators. The energy and coupling parameters can be found so as to improve the quantum memory performance in its storage phase by maximising the memory decoherence time. The latter is  associated with  keeping a mean-square deviation of the system variables or quantum states from their initial conditions below a given  critical threshold. This approach leads to quantum memory optimisation problems,  considered in the coherent settings in [\cite{VP_2024_ANZCC,VP_2024_EJC,VPS_2025_CDC}] and contributing  to the synthesis of coherent memory networks from quantum optical components [\cite{NG_2015}].
The solution of these problems employs the tractability of second and higher-order moment dynamics  for open quantum harmonic oscillators (OQHOs) with positions and momenta  and finite-level open quantum systems with the Pauli matrices [\cite{S_1994}] as dynamic variables. The moment dynamics tractability (at least in the case of input fields in the vacuum state)  holds for these two major classes of open quantum systems  despite the fact that they are governed by qualitatively different  (respectively, linear and quasi-linear) QSDEs.

The present paper is concerned with a quantum memory optimisation setting,  where, in contrast to the coherent case mentioned above, the deviation of the underlying OQHO memory system from its initial condition is reduced by varying the classical parameters of the system Hamiltonian in time through the actuator output of a measurement-based classical controller. The latter is in the form of a linear time-varying (LTV) stochastic system driven by an observation process produced by a measuring apparatus from the quantum output field of the memory system. This  results
in a closed-loop system with both quantum and classical dynamic variables.
We consider a controller with a separation structure [\cite{AM_1989}] which continuously updates  an  estimate for the initial dynamic variables of the memory system. This  initial-point smoother is used for producing a control  signal so as to minimise the sum of a mean-square  deviation of the memory system variables at a given time horizon  from their initial values and an integral quadratic penalty on the actuator signal. To this end, an adaptation of classical linear-quadratic-Gaussian (LQG)  control and fixed-point smoothing [\cite{AM_1979}]  is applied to an appropriately augmented quantum system, with  the quantum and classical random variables being considered at the level of their first two moments.
In comparison with the  stabilisation of quantum states   for multi-qubit systems [\cite{ASDSMR_2013,LAM_2022}], the present approach employs a different class of quantum systems and  classical  controllers along with different optimisation methods.

The paper is organised as follows.
Section~\ref{sec:mem} specifies the class of open quantum memory systems with measurement-based controllers.
Section~\ref{sec:meas} describes classical LTV controllers under  consideration.
Section~\ref{sec:est} considers a controller whose internal state is a mean-square optimal smoothing estimate for the initial memory condition.
Section~\ref{sec:cont} describes the optimal control law in the sense of minimising the terminal-integral quadratic cost functional.
Section~\ref{sec:conc} provides concluding remarks.

\section{Quantum memory with measurement-based feedback}
\label{sec:mem}

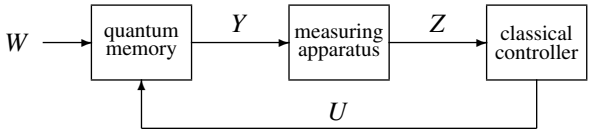
\begin{figure}[htbp]
\unitlength=0.65mm
\linethickness{0.5pt}
\qquad\begin{picture}(110.00,28.00)
    \put(13,10){\framebox(20,15)[cc]{{}}}
    \put(13,12){\makebox(20,15)[cc]{{\small quantum}}}
    \put(13,8){\makebox(20,15)[cc]{{\small memory}}}
    \put(53,10){\framebox(20,15)[cc]{{}}}
    \put(53,12){\makebox(20,15)[cc]{{\small measuring}}}
    \put(53,8){\makebox(20,15)[cc]{{\small apparatus}}}
    \put(93,10){\framebox(20,15)[cc]{{}}}
    \put(93,12){\makebox(20,15)[cc]{{\small classical}}}
    \put(93,8){\makebox(20,15)[cc]{{\small controller}}}

    \put(-2,17.5){\makebox(0,0)[cc]{$W$}}
    \put(43,21.5){\makebox(0,0)[cc]{$Y$}}
    \put(83,21.5){\makebox(0,0)[cc]{$Z$}}
    \put(63,3.5){\makebox(0,0)[cc]{$U$}}
    \put(3,17.5){\vector(1,0){10}}
    \put(33,17.5){\vector(1,0){20}}
    \put(73,17.5){\vector(1,0){20}}
    \put(23,0){\vector(0,1){10}}
    \put(103,0){\line(-1,0){80}}
    \put(103,0){\line(0,1){10}}
\end{picture}
\caption{The quantum memory system is subject to the quantum Wiener process $W$ as environmental noise and also driven by the actuator signal $U$ of a classical controller based on the observation  process $Z$ produced by a measuring apparatus from the output quantum field $Y$ of the system.}
\label{fig:OQHO}
\end{figure}

We consider an open quantum  memory system, further referred to as the plant,  which has internal dynamic variables $X_1, \ldots, X_n$ assembled into a vector $X := (X_k)_{1\< k \< n}$ and organised as time-varying self-adjoint operators (the time argument is often omitted) on a Hilbert space $\fH$ specified below. The plant interacts with external bosonic fields modelled by a quantum Wiener process $W:= (W_k)_{1\< k \< m}$ of an even number $m$ of time-varying self-adjoint operators $W_1, \ldots, W_m$ on a symmetric Fock space $\fF$ [\cite{P_1992,PS_1972}]. These fields are assumed to be in the vacuum state $\ups$ on $\fF$. The plant-field  interaction produces an output quantum field $Y:= (Y_k)_{1\< k \< m}$ consisting of time-varying self-adjoint operators on the plant-field tensor-product   Hilbert space $\fH := \fH_0 \ox \fF$, where $\fH_0$ is the initial plant space for the action of $X_0:= X(0)$. The plant output $Y$   is registered by a measuring apparatus (see Fig.~\ref{fig:OQHO})
and converted to a  multichannel observation process $Z:=(Z_k)_{1\< k\< r}$
consisting of $r\< \frac{m}{2}$ time-varying self-adjoint operators on $\fH$, which commute with each other at all times and   with the plant variables at the same and future moments of time:
\begin{equation}
\label{ZZXcomm}
    [Z(t), Z(s)^\rT]= 0,
    \quad
    [X(t), Z(s)^\rT] = 0,
    \qquad
    t \> s \> 0,
\end{equation}
where $[\alpha,\beta]:= ([\alpha_j, \beta_k])_{1\< j \< a, 1 \< k \< b}
= 
-[\beta,\alpha^\rT]^\rT$ is the matrix of pairwise commutators $[\alpha_j, \beta_k]:= \alpha_j \beta_k - \beta_k \alpha_j$ between linear operators assembled into column-vectors $\alpha := (\alpha_j)_{1\< j \< a}$ and $\beta  := (\beta_k)_{1\< k \< b}$, with $(\cdot)^\rT$ the usual transpose.
Accordingly, $Z$ by itself can be regarded as a classical $\mR^r$-valued random process. It   is fed into a measurement-based classical controller which produces  an $\mR^d$-valued  actuator signal $U:= (U_k)_{1\< k \< d}$. The latter affects the quantum plant dynamics,  thus closing the feedback loop in Fig.~\ref{fig:OQHO} as described below.

The plant is assumed to be an OQHO with an even number $n$  of dynamic variables, which are  organised as pairs of conjugate  quantum mechanical positions and momenta [\cite{S_1994}]) satisfying the one-point canonical commutation relations (CCRs) \begin{equation}
\label{XCCR}
    [X, X^{\rT}]
       =
     2i \Theta,
     \qquad
     \Theta := \frac{1}{2} I_{n/2} \ox \bJ,
     \qquad
            \bJ
       :=
       {\begin{bmatrix}
         0 & 1\\
         -1 & 0
       \end{bmatrix}},
\end{equation}
where $i:= \sqrt{-1}$ is the imaginary unit, $\ox$ is the Kronecker product of matrices, and $I_s$ is the identity matrix of order $s$. The energetics of the plant and its interaction with the external input field $W$ is  specified by the Hamiltonian $H$ and the vector $L:= (L_k)_{1\< k \< m}$ of plant-field coupling operators. These energy and coupling operators are quadratic and linear functions of the plant variables, respectively:
\begin{equation}
\label{HL}
    H
    :=
    \frac{1}{2}
    X^\rT R X
    +
    U^\rT N X,
    \qquad
    L:=
    (L_k)_{1\< k\< m}
    =
    MX,
\end{equation}
where $R = R^{\rT} \in \mR^{n\x n}$ is the energy matrix, and $M \in \mR^{m\x n}$ is the plant-field coupling  matrix. Also, $N \in \mR^{d \x n}$ is the plant-controller coupling  matrix whose rows $N_j \in \mR^{1\x n}$ give rise to the control Hamiltonians
$H_j := N_j X$, with $j = 1, \ldots, d$. Their linear combination $U^\rT NX = \sum_{j = 1}^d U_j H_j$   in (\ref{HL}) captures the effect of the controller actuator output $U$ as a time-varying $\mR^d$-valued classical  parameter of the Hamiltonian $H$. The classicality of $U$ as a parameter of $H$ is justified by the nondemolition condition [\cite{B_1983}] with respect to the plant variables:
\begin{equation}
\label{XUcomm}
  [X(t), U(s)^\rT]
  =
  0,
  \qquad
  t\> s\> 0,
\end{equation}
which is a consequence of the second commutativity from (\ref{ZZXcomm}) and the adaptedness of $U$ to the natural filtration of the observation process $Z$.  Due to the CCRs (\ref{XCCR}) and the linear-quadratic energetics (\ref{HL}), the plant evolves as a linear quantum stochastic system governed by the QSDEs
\begin{equation}
\label{dXdY}
    \rd X
    =
    (A X +  E U) \rd t+ B \rd W,
    \qquad
\rd Y = 2J MX \rd t + \rd W.
\end{equation}
Here, similarly to [\cite{VJP_2020}],  the matrices
$A\in \mR^{n\x n}$, $B \in \mR^{n\x m}$, $E\in \mR^{n\x d}$  are parameterised as
\begin{equation}
\label{ABE}
  A = 2\Theta (R + M^\rT J M),
  \qquad
  B = 2\Theta M^{\rT},
  \qquad
  E = 2\Theta N^\rT
\end{equation}
by the commutation, energy and coupling matrices. In particular, $E$ in (\ref{ABE}) comes from the control part of the Hamiltonian $H$ in (\ref{HL}) as $i[U^\rT N X, X] = -i [X,X^\rT]N^\rT U = E U$ due to (\ref{XCCR}), (\ref{XUcomm}).
Also,
\begin{equation}
\label{J}
    J
        :=
        \Im \Omega
        =
        I_{m/2} \ox \bJ
\end{equation}
is the imaginary part of the quantum Ito matrix $0 \preccurlyeq\Omega = \Omega^* \in \mC^{m \x m}$,
\begin{equation}
\label{WW}
    \rd W \rd W^{\rT}
    =
    \Omega \rd t,
    \qquad
    \Omega := I_m + iJ,
\end{equation}
for the quantum Wiener process $W$ and specifies its two-point CCRs
\begin{equation*}
\label{WWcomm}
    [W(s), W(t)^\rT]
    =
    2i\min(s,t) J,
    \qquad
    s,t\>0.
\end{equation*}
The CCR matrix $J$ is inherited by the plant output field $Y$ in (\ref{dXdY}):
$    [\rd Y, \rd Y^\rT] = 2i J \rd t
$,
whereby $Y_1, \ldots, Y_m$ do not commute with each other and hence, are not accessible to simultaneous measurement.

Note that the first equality in (\ref{HL}) is not the only way to introduce the classical actuator signal $U$ in the Hamiltonian $H$. An alternative is provided by a control-dependent energy matrix $R = R_0 + \sum_{k=1}^d U_k R_k$,  where $R_0, \ldots, R_d$ are real symmetric matrices of order $n$. In this case, the role of control Hamiltonians is played by the operators  $\frac{1}{2} X^\rT R_k X$, with $k = 1, \ldots, d$, and the first QSDE in (\ref{dXdY}) is replaced with $\rd X
    =
    (A_0 + \sum_{k=1}^d U_k A_k) X \rd t+ B \rd W$ (with the matrices $A_k:= 2\Theta R_k$ for all $k = 0, \ldots, d$), which involves the actuator signal $U$ in a multiplicative, rather than linear, fashion. However, the multiplicativity makes the system dynamics nonlinear and the resulting optimisation problems less tractable. In addition to the above ways of influencing the quantum plant through its Hamiltonian, the control can also act as an additional  field-like term (not dissimilar to $B\rd W$ in (\ref{dXdY})), which will be discussed elsewhere.

\section{Linear time-varying controllers with nondemolition measurements}
\label{sec:meas}

Similarly to [\cite{N_2014}], the measuring apparatus in Fig.~\ref{fig:OQHO} is modelled by a static linear relation of the classical observation process $Z$ to the plant output $Y$:
\begin{equation}
\label{ZY}
    Z
     = D Y,
\end{equation}
where $D\in \mR^{r\x m}$ is a constant matrix satisfying the conditions
\begin{equation}
\label{DD_DJD}
  G:= DD^{\rT}  \succ 0,
  \qquad
  DJD^{\rT}  = 0,
\end{equation}
where the matrix $J$ is given by (\ref{J}).
The inequality in (\ref{DD_DJD}) is equivalent to $D$ being of full row rank, while the second equality secures the commutativity in (\ref{ZZXcomm}). By the second QSDE in (\ref{dXdY}), the process $Z$ in (\ref{ZY}) satisfies another QSDE
\begin{equation}
\label{dZ}
  \rd Z = CX \rd t + D \rd W,
\end{equation}
where the matrix $C\in \mR^{r\x n}$ depends on the plant-field coupling matrix $M$ from (\ref{HL}) as
\begin{equation}
\label{C}
  C:= 2DJM.
\end{equation}
The process $DW$, which drives (\ref{dZ}), has the Ito matrix $D\Omega D^\rT = G$ in view of (\ref{WW}), (\ref{DD_DJD}), and  is isomorphic to a classical Wiener process in $\mR^r$ with a nonsingular diffusion matrix $G$ which is also shared by $Z$ as a classical Ito process. The latter drives the controller which is assumed to be in the form of a classical LTV system with an $\mR^\nu$-valued state process $x := (x _k)_{1\< k \<  \nu}$  described by the following  Ito SDE and linear output relation:
\begin{equation}
\label{dxU}
  \rd x   = a x  \rd t + b \rd Z,
  \qquad
  U  = c x,
\end{equation}
where the initial condition $x (0)$ is $\fH_0$-adapted,  and $a$, $b$, $c$ are arbitrary continuous functions of time with values in
$\mR^{\nu\x \nu}$, $\mR^{\nu\x r}$, $\mR^{d\x \nu}$, respectively (the controller state dimension $\nu$ is specified in the next section).
The closed-loop
system acquires an augmented vector
\begin{equation}
\label{cX}
  \cX
  :=
  {\begin{bmatrix}
    X\\
    x
  \end{bmatrix}},
\end{equation}
formed from the dynamic variables of the quantum plant (\ref{dXdY}) and the classical controller (\ref{dxU}). Regardless of a particular control law (for $U$ as a function of the past history of the observation process $Z$) in (\ref{dxU}), the process $\cX$ in (\ref{cX}) satisfies the one-point CCRs
\begin{equation*}
\label{cXCCR}
    [\cX,\cX^\rT]
  =
    2i
  {\begin{bmatrix}
    \Theta & 0\\
    0 & 0
  \end{bmatrix}}
\end{equation*}
in view of (\ref{XCCR}) and the classical nature of the controller variables. However, the linear control law in (\ref{dxU}) makes $\cX$ evolve according to a linear QSDE
\begin{equation*}
\label{dcX}
  \rd \cX = \cA \cX \rd t + \cB \rd W.
\end{equation*}
Here, $\cA$, $\cB$ are time-varying matrices which take values in $\mR^{(n+\nu)\x (n+\nu)}$ and  $\mR^{(n+\nu)\x m}$, respectively, and are computed as
\begin{equation*}
\label{cAB}
  \cA
  :=
  {\begin{bmatrix}
    A & Ec\\
    bC & a
  \end{bmatrix}},
  \qquad
  \cB
  :=
  {\begin{bmatrix}
    B\\
    bD
  \end{bmatrix}},
\end{equation*}
thus inheriting the  time dependence from the controller matrices $a$, $b$, $c$. The latter can be varied so as to improve the ability of the plant (as part of the closed-loop memory system) to approximately retain the initial condition $\varphi(0)$ of an auxiliary quantum process
\begin{equation}
\label{FX}
    \varphi:= FX,
\end{equation}
consisting of $s\< n$ time-varying self-adjoint operators on $\fH$ and satisfying the QSDE
\begin{equation}
\label{dphi}
  \rd \varphi = F(A X +  E U) \rd t+ FB \rd W
\end{equation}
in view of the first QSDE from (\ref{dXdY}).
Here,
$F \in \mR^{s \x n}$ is a given full row rank matrix which specifies independent linear combinations (for example, a subset) of the plant variables of interest. The memory performance can then be quantified in terms of a mean-square deviation functional [\cite{VP_2024_ANZCC,VP_2024_EJC}]
\begin{equation}
\label{Del}
    \Delta(t) := \bE (\eta(t)^\rT\eta(t)) = \bE (\xi(t)^\rT \Sigma \xi(t)).
\end{equation}
Here, $\bE \zeta := \Tr(\rho \zeta)$ is the quantum expectation over the density operator $\rho = \rho_0 \ox \ups$ on the plant-field space $\fH$, with $\rho_0$ the initial plant state on $\fH_0$. Also,
\begin{equation}
\label{xieta}
    \xi(t):= X(t)-X_0,
    \qquad
    \eta(t):= \varphi(t) - \varphi(0) = F \xi(t),
\end{equation}
in accordance with (\ref{FX}),
and use is made of a real positive semi-definite symmetric matrix
\begin{equation}
\label{FF}
    \Sigma := F^\rT F
\end{equation}
of order $n$
satisfying     $\rank \Sigma = s$. Accordingly, the mean-square optimisation of the quantum memory system is related to the minimisation of (\ref{Del})
at a given time horizon (and the actuator signal $U$  over the time interval)  with respect to the controller matrices $a$, $b$, $c$ as functions of time.

\section{Mean-square optimal smoothing for initial memory condition}
\label{sec:est}

At every moment of time, the processes  $\xi$, $\eta$ in (\ref{xieta}) are related by
\begin{equation}
\label{xietasX}
  \xi = (\begin{bmatrix}
     -1 & 1
  \end{bmatrix}\ox I_n)\sX,
  \qquad
  \eta = (\begin{bmatrix}
     -1 & 1
  \end{bmatrix}
  \ox F)\sX
\end{equation}
to a quantum process $\sX$ defined by augmenting the current plant variables  by their initial values:
\begin{equation}
\label{XX}
  \sX(t)
  :=
  \begin{bmatrix}
    X_0\\
    X(t)
  \end{bmatrix},
  \qquad
  t\> 0.
\end{equation}
The dynamics of $\sX$ and the observation process $Z$ are governed by appropriately augmented versions of the QSDEs (\ref{dXdY}), (\ref{dZ}):
\begin{equation}
\label{dsXZ}
  \rd \sX
  =
  (\sA
  \sX + \sE U)
  \rd t
  +
  \sB
  \rd W,
  \qquad
  \rd Z =
  \sC
  \sX \rd t + D \rd W,
\end{equation}
with
the matrices $\sA \in \mR^{2n\x 2n}$, $\sB \in \mR^{2n \x m}$, $\sC\in \mR^{r\x 2n}$, $\sE\in \mR^{2n\x d}$ given by
\begin{equation}
\label{sABCE}
  \sA :=   \begin{bmatrix}
    0 & 0\\
    0 & A
  \end{bmatrix},
  \quad
  \sB:=
  \begin{bmatrix}
    0 \\
    B
  \end{bmatrix},
  \quad
  \sC
  :=
  \begin{bmatrix}
    0 & C
  \end{bmatrix},
  \quad
  \sE:=
  \begin{bmatrix}
    0 \\
    E
  \end{bmatrix}
\end{equation}
in terms of (\ref{ABE}), (\ref{C}). The sparsity of the matrices (\ref{sABCE}) comes from the fact that $X_0$ remains constant in time.
Similarly to the classical fixed-point smoothing techniques (which are discussed, for example, in \cite[Section 7.2]{AM_1979} in the discrete-time case), we consider a controller (\ref{dxU}) with the state dimension $\nu := 2n$ and the following internal dynamics:
\begin{equation}
\label{dxUK}
    \rd x
     =
    (\sA x + \sE U)  \rd t + K\rd V,
\end{equation}
initialised at
\begin{equation}
\label{x0}
  x(0)
  :=
  \bE \sX(0)
  =
  \begin{bmatrix}
    1\\
    1
  \end{bmatrix}
  \ox
  \bE X_0
\end{equation}
in view of (\ref{XX}). The equation (\ref{dxUK}) is a classical SDE which is driven by an $\mR^r$-valued Ito process $V$
playing  the role of an innovation process whose increments
\begin{equation}
\label{dV}
  \rd V
  :=
  \rd Z - \sC x\rd t
  =
  \sC e\rd t + D \rd W
\end{equation}
are weighted in (\ref{dxUK})
by an $\mR^{2n\x r}$-valued continuous function of time $K$, so that the matrices $a$, $b$ in (\ref{dxU}) are given  by
\begin{equation}
\label{abK}
  a = \sA + \sE c - K \sC,
  \qquad
  b = K.
\end{equation}
In (\ref{dV}),  use is made of  the second QSDE from  (\ref{dsXZ}) along with the error process
\begin{equation}
\label{e}
  e:= \sX - x,
\end{equation}
which consists of $2n$ time-varying self-adjoint operators satisfying
the QSDE
\begin{equation}
\label{de}
  \rd e
  =
  (\sA-K\sC) e \rd t+  (\sB-KD) \rd W.
\end{equation}
The latter is obtained by subtracting the SDE  (\ref{dxUK}) from the first QSDE in (\ref{dsXZ}), similarly to the  derivations in the context of the classical Luenberger observer (including the Kalman filter). 
Since the external fields are assumed to be in the vacuum state,  the averaging of (\ref{de}) leads to $(\bE e)^{^\centerdot} = (\sA-K\sC )\bE e$, where $\dot{(\ )}:= \frac{\rd}{\rd t}$ is the time  derivative. The last ODE, combined with the initial condition $\bE e(0) = \bE \sX(0) - x(0)=0$ from (\ref{XX}), (\ref{x0}), implies that the error process retains zero mean value over the course of time:
\begin{equation}
\label{Ee0}
    \bE e(t) = 0,
    \qquad
    t \in [0,\tau].
\end{equation}
Furthermore, from (\ref{WW}), (\ref{de}), (\ref{Ee0}),  it follows that the real part
\begin{equation}
\label{P}
  P:= \Re \cov(e)
  =
    \begin{bmatrix}
      P_1 & P_2\\
      P_2^\rT & P_3
    \end{bmatrix}
\end{equation}
(partitioned into square blocks of order $n$) 
of the one-point quantum covariance matrix $\cov(e):= \bE(ee^\rT) - \bE e \bE e^\rT = \bE(ee^\rT)$ satisfies a Lyapunov ODE
\begin{align}
\nonumber
    \dot{P}
    & =
    (\sA-K\sC) P + P (\sA-K\sC)^\rT + (\sB-KD)(\sB-KD)^\rT \\
\nonumber
    & =
    \sA P + P \sA^\rT + \sB\sB^\rT -
    K_* G K_*^\rT+
    (K-K_*)G(K-K_*)^\rT\\
\label{Pdot}
    & \succcurlyeq
\sA P + P \sA^\rT + \sB\sB^\rT -
    K_* G K_*^\rT,
\end{align}
where $G$ is the observation diffusion matrix from (\ref{DD_DJD}). The minimal (in the sense of the partial ordering induced by positive semi-definiteness) solution $P$ of (\ref{Pdot}) is delivered by the Kalman gain matrix
\begin{equation}
\label{KK*}
    K =
    K_*:=
    (P \sC^\rT + \sB D^\rT)
    G^{-1}
    =
    \begin{bmatrix}
      P_2 C^\rT\\
      P_3 C^\rT + BD^\rT
    \end{bmatrix}
    G^{-1}.
\end{equation}
In this case, (\ref{Pdot})  takes the form of the Kalman filter  Riccati ODE in application to (\ref{dsXZ}):
\begin{equation}
\label{PRic}
    \dot{P}
    =
    \sA P + P \sA^\rT + \sB\sB^\rT -
    K G K^\rT,
\end{equation}
with the controller variables and the error process being uncorrelated in the sense that
\begin{equation}
\label{Exe}
    \mho  := \Re \bE (xe^\rT) = 0.
\end{equation}
The latter property complements (\ref{Ee0}) and follows from the Sylvester ODE
\begin{align*}
    \dot{\mho}
    & =
    (\sA + \sE c) \mho + K\sC P + \mho (\sA- K\sC)^\rT
    + KD (B-KD)^\rT \\
    & =
    (\sA + \sE c) \mho + \mho (\sA- K\sC)^\rT
    + K(\sC P + DB^\rT -G K^\rT) \\
    & =
    (\sA + \sE c) \mho + \mho (\sA- K\sC)^\rT,
\end{align*}
obtained from (\ref{dxUK}), (\ref{dV}), (\ref{de}), (\ref{KK*}) and
initialised at $\mho(0) = x(0) \Re \bE e(0)^\rT= 0$.
The partitioning in (\ref{P}), (\ref{KK*}) allows (\ref{PRic}) to be represented as a set of three ODEs
\begin{align}
\label{P1dot}
    \dot{P}_1
    & =
    - P_2 C^\rT G^{-1} C P_2^\rT , \\
\label{P2dot}
    \dot{P}_2
    & =
    P_2
    (A^\rT - C^\rT G^{-1}(CP_3 + DB^\rT)), \\
\label{P3dot}
    \dot{P}_3
    & =  \!
    A P_3 \!+\! P_3 A^\rT \!+\! BB^\rT \!-\! (P_3 C^\rT  \!+\! BD^\rT)
    G^{-1} (CP_3 \!+\! DB^\rT)
\end{align}
initialised at the real part of the covariance matrix of the initial plant variables:
\begin{equation}
\label{P1230}
    P_k(0) = \Re \cov(X_0),
    \qquad
    k = 1, 2, 3.
\end{equation}
While (\ref{P3dot}) is an autonomous Riccati ODE for the matrix $P_3$, the ODE (\ref{P1dot}) for $P_1$ is driven by $P_2$, which in turn is driven by $P_3$ according to (\ref{P2dot}).

The controller state vector $x$ in (\ref{dxUK}) with the gain matrix (\ref{KK*}) provides a mean-square optimal estimate of the quantum vector $\sX$ in (\ref{XX}). More precisely, this optimality  is in the sense of minimising the real part $\Re \bE (ee^\rT)$ of the second-moment matrix for the error process (\ref{e}) in the class of linear observers governed by the SDE from (\ref{dxU}). Accordingly, the process $x$ is partitioned as
\begin{equation}
\label{xXX}
   x(t)
   =
   \begin{bmatrix}
     \wh{X}_0(t)\\
     \wh{X}(t)
   \end{bmatrix},
   \qquad
   t \> 0,
\end{equation}
where the $n$-dimensional subvectors $\wh{X}_0(t)$, $\wh{X}(t)$ are the smoothing and filtering estimates of $X_0$ and $X(t)$, respectively,  based on the past history
\begin{equation}
\label{cZt}
    \cZ_t:= Z|_{[0,t]}
\end{equation}
of the observation process $Z$ available at time $t$. Therefore, the controller internal dynamics (\ref{dxUK}) acquire the form of two SDEs
\begin{align}
\label{dX0hat}
    \rd \wh{X}_0
    & =
    P_2 C^\rT G^{-1}
    \rd V,\\
\label{dXhat}
    \rd \wh{X}
    & =
    (A \wh{X}  + EU)\rd t
    +
    (P_3 C^\rT + BD^\rT )
    G^{-1}
    \rd V,
\end{align}
which take into account the block structure of the matrices
(\ref{sABCE}), (\ref{KK*}) and share a common initial condition
\begin{equation}
\label{XX0}
  \wh{X}_0(0) = \wh{X}(0) = \bE X_0
\end{equation}
in accordance with (\ref{x0}).
Here, the stochastic   differential $\rd V$ in   (\ref{dV}) involves $\wh{X}$ from  (\ref{xXX}) as
\begin{equation}
\label{dVX}
  \rd V
  =
  \rd Z - C \wh{X}\rd t.
\end{equation}
Therefore, the filtering estimate $\wh{X}$ enters the right-hand side of the SDE (\ref{dX0hat}) for updating the smoothing estimate $\wh{X}_0$ through the process $V$.

If (\ref{dsXZ}) were usual SDEs in Euclidean spaces, driven by a standard Wiener process $W$     and initialised at a Gaussian distributed $X_0$ with the covariance matrix (\ref{P1230}) and mean (\ref{XX0}), then,  at any time $t\> 0$, the vector $x(t)$ in (\ref{xXX}) would coincide with the classical conditional expectation $\bM(\sX(t) \mid \cZ_t)$ of $\sX(t)$ given the observation history (\ref{cZt}).

Also note that the solution $P$ of the Riccati ODE (\ref{PRic}) (and its infinite-horizon behaviour, as $t\to +\infty$) can be represented in terms of the
Hamiltonian matrix
\begin{equation}
\label{sH}
    \sH:=
    \begin{bmatrix}
      \alpha & \beta\\
      \gamma & - \alpha^\rT
    \end{bmatrix}
    =
    (\bJ \ox I_{2n})
    \begin{bmatrix}
      -\gamma & \alpha^\rT\\
      \alpha &  \beta
    \end{bmatrix},
\end{equation}
where $\alpha$, $\beta = \beta^\rT$ and $\gamma = \gamma^\rT$ are real matrices of order $2n$ given by
\begin{align*}
    \alpha
    & := \sA - \sB D^\rT G^{-1} \sC
    =
    \begin{bmatrix}
      0 & 0 \\
      0 & A - B D^\rT G^{-1} C
    \end{bmatrix}, \\
    \beta
    & := \sB (I_m - D^\rT G^{-1} D)\sB^\rT
    =
    \begin{bmatrix}
      0 & 0 \\
      0 &  B (I_m - D^\rT G^{-1} D)B^\rT
    \end{bmatrix}, \\
    \gamma
    & := \sC^\rT G^{-1} \sC
    =
    \begin{bmatrix}
      0 & 0 \\
      0 &  C^\rT G^{-1} C
    \end{bmatrix}.
\end{align*}
However, in contrast to the usual assumption (of no eigenvalues on the imaginary axis), the matrix $\sH$ in (\ref{sH}) is singular  because it is isospectral to the matrix
$$
    \begin{bmatrix}
      0 & 0 \\
      0 & 1
    \end{bmatrix}
    \ox
    \begin{bmatrix}
      A - B D^\rT G^{-1} C & & B (I_m - D^\rT G^{-1} D)B^\rT\\
      C^\rT G^{-1} C & & -(A - B D^\rT G^{-1} C)^\rT
    \end{bmatrix}.
$$
The structure of the nontrivial null space $\ker \sH$ and its effect on
the behaviour of the solution $P$ of (\ref{PRic})
will be discussed elsewhere.

\section{Finite-horizon mean-square optimal memory control}
\label{sec:cont}

The closed-loop memory performance at a given time horizon $\tau>0$ can be optimised in the  sense of minimising a terminal-integral quadratic cost:
\begin{align}
\nonumber
    \Phi(\tau)
    & :=
  \Delta(\tau) + \int_0^\tau \bE (\|U(t)\|_\Pi^2) \rd t\\
\label{cost}
  & = \bE \Big(\sX(\tau)^\rT \Lambda \sX(\tau) + \int_0^\tau \|U(t)\|_\Pi^2 \rd t\Big)
  \longrightarrow \inf,
\end{align}
where use is made of the mean-square deviation functional (\ref{Del}) represented  as
\begin{equation}
\label{DelX}
  \Delta(\tau)
  =
  \bE (\sX(\tau)^\rT \Lambda \sX(\tau)),
  \qquad
  \Lambda
  :=
  \begin{bmatrix}
    1& -1\\
    -1 & 1
  \end{bmatrix}
  \ox \Sigma
\end{equation}
in terms of (\ref{FF}), (\ref{XX}) in view of (\ref{xietasX}). Recall that $\Delta(\tau)$ quantifies the accuracy with which the quantum plant variables of interest, captured in the process (\ref{FX}), approximately reproduce at time $\tau$ their initial values. The minimisation in (\ref{cost}) is over the time-varying feedback gain matrix $c\in C([0,\tau], \mR^{d \x 2n})$ of the controller described by (\ref{dxU}), (\ref{dxUK}), (\ref{abK}), (\ref{P}), (\ref{KK*})--(\ref{dVX}).
Here, $\|u\|_\Pi : = \sqrt{u^\rT \Pi u} = |\sqrt{\Pi}u|$ is a weighted Euclidean norm of a vector $u \in \mR^d$, and $0\prec \Pi = \Pi^\rT \in \mR^{d\x d}$ is  a given matrix which specifies the quadratic penalty on the actuator signal $U$. Moderate values of $U$ can be preferable because of physical setup limitations or in order to avoid nonlinear effects in the system dynamics. Alternatively, the integral term in (\ref{cost}) can also play the role of  a Tikhonov regularisation for making $\Phi(\tau)$  a strongly convex functional of $U$.

The first equality in (\ref{DelX}) and the symmetry of the matrix $\Lambda$   allow the terminal mean-square deviation to be represented as
\begin{equation}
\label{DelS}
  \Delta(\tau)
  =
  \bra
    \Lambda,
    S(\tau)
  \ket ,
\end{equation}
where $\bra \cdot, \cdot \ket$ is the Frobenius inner product of matrices.
Here, the real part
\begin{align}
\nonumber
  S
  & :=
  \Re
     \bE (\sX \sX^\rT)
     =
  \Re
     \bE ((x+e) (x+e)^\rT)\\
\label{S}
     & =
     T + \mho + \mho^\rT + P = T+P
\end{align}
of the second-moment matrix of the quantum process $\sX$ from (\ref{XX}) is expressed in terms of (\ref{e}), (\ref{P}), (\ref{Exe}) and the the second-moment matrix
\begin{equation}
\label{T}
    T:= \bE (xx^\rT)
\end{equation}
of the controller variables. Accordingly, the integrand  in (\ref{cost}) takes the form
\begin{equation}
\label{EU2}
  \bE (\|U\|_\Pi^2) = \bra c^\rT \Pi c, T\ket
\end{equation}
using the second equality from (\ref{dxU}). By (\ref{dxUK}), (\ref{dV}), the matrix (\ref{T})
satisfies a Lyapunov ODE
\begin{equation}
\label{Tdot}
    \dot{T} = (\sA + \sE c)T + T (\sA + \sE c)^\rT + KGK^\rT =: \cR(T,c)
\end{equation}
with the initial condition
\begin{equation}
\label{T0}
    T(0) =
    \begin{bmatrix}
    1 & 1\\
    1 & 1
  \end{bmatrix}
  \ox
  (\bE X_0 \bE X_0^\rT).
\end{equation}
in view of (\ref{x0}). Since the diffusion matrix $G$ is constant, and the Kalman gain matrix $K$ in  (\ref{KK*}) and the real part $P$ of the error covariance matrix in (\ref{P}) are independent of the control law, the relations (\ref{DelS}), (\ref{S}), (\ref{EU2}) reduce the optimisation problem in (\ref{cost}) to that of minimising
\begin{equation}
\label{redcost}
    \bra \Lambda, T(\tau)\ket + \int_0^\tau \bra c^\rT \Pi c, T\ket\rd t
    \longrightarrow
    \inf
\end{equation}
over the controller gain matrix $c\in C([0,\tau], \mR^{d \x 2n})$ which plays the role of control on the right-hand side of (\ref{Tdot}).
Similarly to the structure of the optimal controller in classical LQG  control [\cite{AM_1989}] (including  the separation principle), the minimum in (\ref{cost}) is delivered by the feedback gain matrix
\begin{equation}
\label{copt}
  c = -\Pi^{-1} \sE^\rT Q = -\Pi^{-1} E^\rT \begin{bmatrix}
    Q_2 & Q_3
  \end{bmatrix}
\end{equation}
on the time interval $[0,\tau]$,
where use is made of the matrix $\sE$ from (\ref{sABCE}). Here, $Q$ is a time-varying real positive semi-definite symmetric matrix,   which is partitioned into square blocks of order $n$ as
\begin{equation}
\label{Q}
    Q:=
      \begin{bmatrix}
      Q_1 & Q_2^\rT\\
      Q_2 & Q_3
    \end{bmatrix}
\end{equation}
and is the solution of the boundary value problem for the control Riccati ODE
\begin{equation}
\label{QRic}
  \dot{Q}
  =
  Q \sE \Pi^{-1} \sE^\rT Q
  -\sA^\rT Q - Q \sA
\end{equation}
on the time interval $[0,\tau]$,  with the terminal condition $Q(\tau) = \Lambda$ from (\ref{DelX}). The latter is equivalent to
\begin{equation}
\label{Q123fin}
  Q_k(\tau) = -(-1)^k \Sigma,
  \qquad
  k = 1, 2, 3,
\end{equation}
in view of the partitioning (\ref{Q}). The matrix $Q$ specifies the Bellman function for the linear-quadratic control problem (\ref{redcost}) with the dynamics (\ref{Tdot}):
\begin{equation}
\label{Psi}
    \Psi(t,\Gamma)
    =
    \bra Q(t), \Gamma\ket
    +
    \int_t^\tau
    \bra Q(v), K(v)GK(v)^\rT\ket \rd v
\end{equation}
for all     $t \in [0,\tau]$ and $\Gamma \in \mS_{2n}^+$, with
$\mS_r^+$ the set of real positive semi-definite symmetric matrices of order $r$.  The function $\Psi$ satisfies the corresponding Hamilton-Jacobi-Bellman equation
\begin{equation}
\label{HJBE}
    \d_t \Psi + \min_{u \in \mR^{d\x 2n}}
    (\bra
        \d_\Gamma \Psi,
        \cR(\Gamma, u)
    \ket  + \bra u^\rT \Pi u, \Gamma\ket) = 0
\end{equation}
(with the boundary condition $\Psi(\tau, \Gamma) = \bra \Lambda, \Gamma\ket$),
where  the Frechet derivative of (\ref{Psi}) with respect to $\Gamma$ takes the form $\d_\Gamma \Psi = Q$. The matrix $u = c$ from (\ref{copt})  delivers the minimum in (\ref{HJBE}) for any $\Gamma \in \mS_{2n}^+$ (the positive semi-definiteness of $\Gamma$ is essential). On the resulting optimal trajectory of (\ref{Tdot}),  the minimum value in (\ref{HJBE}) coincides with that of the Pontryagin control Hamiltonian [\cite{PBGM_1962,SW_1997}]
\begin{align}
\nonumber
    \cH
    & :=
    \bra
        Q,
        \cR(T, c)
    \ket  + \bra c^\rT \Pi c, T\ket\\
\nonumber
    & =
    \bra
        (\sA + \sE c)^\rT Q + Q(\sA + \sE c) + c^\rT \Pi c,
        T
    \ket
    +
    \bra Q, KGK^\rT\ket\\
\label{cH}
    & =
    \bra Q, KGK^\rT\ket - \bra \dot{Q}, T\ket
\end{align}
which remains constant over the time interval $[0,\tau]$.  The last equality in (\ref{cH}) is obtained by combining (\ref{QRic}) with (\ref{copt}).

The partitioning in (\ref{Q}) allows (\ref{QRic}) to be represented in the block-wise form
\begin{align}
\label{Q1dot}
    \dot{Q}_1
    & =
    Q_2^\rT E \Pi^{-1} E^\rT Q_2,\\
\label{Q2dot}
    \dot{Q}_2
    & =
    (Q_3 E \Pi^{-1} E^\rT - A^\rT) Q_2,\\
\label{Q3dot}
    \dot{Q}_3
    & =
    Q_3 E \Pi^{-1} E^\rT Q_3 - A^\rT Q_3 - Q_3 A.
\end{align}
Similarly to (\ref{P1dot})--(\ref{P3dot}), the ODEs (\ref{Q1dot})--(\ref{Q3dot}) are coupled in a cascade fashion, so that (\ref{Q1dot}) for $Q_1$  is driven by $Q_2$, and (\ref{Q2dot}) for $Q_2$ is driven by $Q_3$, while the latter has the autonomous dynamics (\ref{Q3dot}). By substituting (\ref{copt}) into the second equality in (\ref{dxU}) and using (\ref{xXX}), it follows that the optimal actuator signal takes the form
\begin{equation}
\label{Uopt}
    U
   =
   -\Pi^{-1} E^\rT (Q_2 \wh{X}_0 + Q_3 \wh{X}).
\end{equation}
The right-hand side of (\ref{Uopt}) involves both the smoothing $\wh{X}_0$ and filtering $\wh{X}$ estimates of the initial and current quantum  plant variables,  produced as the subvectors of the controller variables according to the SDEs (\ref{dX0hat}), (\ref{dXhat}).

The role of the control law (\ref{Uopt}) in steering the process $\varphi$ in  (\ref{FX})  back to its initial condition $\varphi_0:= \varphi(0)$ can be directly seen at the time horizon $\tau$, when  it takes the form
\begin{align}
\nonumber
    U(\tau)
    & = -\Pi^{-1} E^\rT \Sigma (\wh{X}(\tau)-\wh{X}_0(\tau))\\
\label{Utau}
    & =
    -\Pi^{-1} E^\rT F^\rT (\wh{\varphi}(\tau)-\wh{\varphi}_0(\tau))
\end{align}
due to the terminal conditions (\ref{Q123fin}) and the structure of the matrix (\ref{FF}), with $\wh{\varphi}:= F \wh{X}$ and $\wh{\varphi}_0:= F \wh{X}_0$ the corresponding estimates. Indeed, the contribution
\begin{equation}
\label{FEUtau}
    FEU(\tau) = -FE \Pi^{-1} E^\rT F^\rT (\wh{\varphi}(\tau)-\wh{\varphi}_0(\tau))
\end{equation}
of (\ref{Utau}) to the drift of the QSDE (\ref{dphi}) is organised
as a negative proportional feedback control with respect to the estimated deviation of $\varphi(\tau)$ from its initial condition  $\varphi_0$ since the gain matrix in (\ref{FEUtau}) satisfies $F E\Pi^{-1} E^\rT F^\rT\succcurlyeq 0$.

The above described controller is optimal in the sense of the minimisation problem (\ref{cost}) among LTV controllers of arbitrary internal dimensions. The minimum value of the cost  can be computed by combining (\ref{DelS}), (\ref{S}) with the Bellman function (\ref{Psi}) as
\begin{align}
\nonumber
    \min\Phi(\tau)
    = &
    \bra
        \Lambda,
        P(\tau)
    \ket
    +
    \Psi(0, T(0))\\
\nonumber
    = &
    \bra
        \Lambda,
        P(\tau)
    \ket + \bra Q(0), T(0)\ket
    +
    \int_0^\tau
    \bra
        Q,
        K G K^\rT
    \ket
    \rd t,
\end{align}
which involves (\ref{T0}) and can be related to the Pontryagin control Hamiltonian (\ref{cH}) whose integral over the time interval $[0,\tau]$ is $\tau \cH$.

Similarly to \cite[Theorem 5]{VPS_2025_IFAC}, there is a link between
the minimisation of the cost $\Phi(\tau)$ in  (\ref{cost}) over LTV controllers at a given time horizon $\tau$  and the maximisation of a modified version of the memory decoherence time
\begin{equation*}
\label{tauPhi}
    \min
    \{
        t\> 0:\
        \Phi(t)
        \>
        \eps
        \Phi_*
    \}
    \longrightarrow \sup,
\end{equation*}
that is, the first moment of time when the cost (\ref{cost})  achieves a critical threshold value $\eps \Phi_*$. The latter is specified by a reference scale constant $\Phi_*>0$ and  a dimensionless fidelity parameter $\eps >0 $. This connection will be discussed elsewhere.

\section{Conclusion}
\label{sec:conc}

We have outlined an application of the classical LQG control and fixed-point smoothing methods (at the level of the first two moments of quantum and classical random processes) to the design of an optimal measurement-based classical LTV controller acting on a quantum memory system through its Hamiltonian. The optimality is in the sense of minimising   the sum of a weighted mean-square deviation of the quantum system variables (to be preserved over the course of time) at a given horizon and an integral  quadratic cost of the control action over the time interval. The controller has a separation structure and involves an initial and boundary value problems for two independent Riccati ODEs and continuously updated mean-square optimal initial point smoothing and filtering estimates for the quantum memory variables,  which are used for linear actuator signal formation.
While the present development is concerned with quantum memory systems in the form of OQHOs, the approach can also be adapted to finite-level quantum systems.

%

\end{document}